\PassOptionsToPackage{table}{xcolor}
\documentclass[manuscript,screen,nonacm]{acmart}

\usepackage{url,hyperref,lineno,microtype,subcaption}
\usepackage{listings}
\usepackage{amsmath, xparse}
\usepackage{soul}
\usepackage{graphicx}
\usepackage{booktabs}
\usepackage{array} 
\usepackage{amsmath}

\usepackage{tcolorbox}

\definecolor{codegreen}{rgb}{0,0.6,0}
\definecolor{codegray}{rgb}{0.5,0.5,0.5}
\definecolor{codepurple}{rgb}{0.58,0,0.82}
\definecolor{backcolour}{rgb}{1,1,1}
\definecolor{lightergray}{gray}{0.9}
\definecolor{lightblue}{rgb}{0.88,0.94,1}

\AtBeginDocument{%
  \providecommand\BibTeX{{%
    \normalfont B\kern-0.5em{\scshape i\kern-0.25em b}\kern-0.8em\TeX}}}

\begin{document}

\title{Discovering Governing Equations of Geomagnetic Storm Dynamics with Symbolic Regression%\\
}

\author{Stefano Markidis}
\email{markidis@kth.se}
\affiliation{%
  \institution{KTH Royal Institute of Technology}
  %\streetaddress{P.O. Box 1212}
  \city{Stockholm}
  %\state{Ohio}
  \country{Sweden}
  %\postcode{43017-6221}
}

\author{Jonah Ekelund}
\email{jonahek@kth.se}
\affiliation{%
  \institution{KTH Royal Institute of Technology}
  %\streetaddress{P.O. Box 1212}
  \city{Stockholm}
  %\state{Ohio}
  \country{Sweden}
  %\postcode{43017-6221}
}

\author{Luca Pennati}
\email{pennati@kth.se}
\affiliation{%
  \institution{KTH Royal Institute of Technology}
  %\streetaddress{P.O. Box 1212}
  \city{Stockholm}
  %\state{Ohio}
  \country{Sweden}
  %\postcode{43017-6221}
}

\author{Andong Hu}
\email{andonghu@kth.se}
\affiliation{%
  \institution{KTH Royal Institute of Technology}
  %\streetaddress{P.O. Box 1212}
  \city{Stockholm}
  %\state{Ohio}
  \country{Sweden}
  %\postcode{43017-6221}
}

\author{Ivy Peng}
\email{ivypeng@kth.se}
\affiliation{%
  \institution{KTH Royal Institute of Technology}
  %\streetaddress{P.O. Box 1212}
  \city{Stockholm}
  %\state{Ohio}
  \country{Sweden}
  %\postcode{43017-6221}
}

\renewcommand{\shortauthors}{S. Markidis et al.}

\begin{abstract}
Geomagnetic storms are large-scale disturbances of the Earth’s magnetosphere driven by solar wind interactions, posing significant risks to space-based and ground-based infrastructure. The Disturbance Storm Time (Dst) index quantifies geomagnetic storm intensity by measuring global magnetic field variations. This study applies symbolic regression to derive data-driven equations describing the temporal evolution of the Dst index. We use historical data from the NASA OMNIweb database, including solar wind density, bulk velocity, convective electric field, dynamic pressure, and magnetic pressure. The PySR framework, an evolutionary algorithm-based symbolic regression library, is used to identify mathematical expressions linking dDst/dt to key solar wind. The resulting models include a hierarchy of complexity levels and enable a comparison with well-established empirical models such as the Burton-McPherron-Russell and O’Brien-McPherron models. The best-performing symbolic regression models demonstrate superior accuracy in most cases, particularly during moderate geomagnetic storms, while maintaining physical interpretability. Performance evaluation on historical storm events includes the 2003 Halloween Storm, the 2015 St. Patrick’s Day Storm, and a 2017 moderate storm. The results provide interpretable, closed-form expressions that capture nonlinear dependencies and thresholding effects in Dst evolution.
\end{abstract}


\maketitle

\section{Introduction}
Space weather investigates the dynamics of the near-Earth space environment driven by solar activity, including solar wind, geomagnetic field disturbances, and energetic particles~\cite{baker1998space}. One of the most significant events in space weather is the occurrence of geomagnetic storms, which are large-scale disturbances of Earth’s magnetosphere caused by increased interactions between the solar wind and the magnetosphere, such as coronal mass ejections. Geomagnetic storms strongly impact human assets in space and on the ground; they can affect satellite operations, navigation systems, power grids, and high-frequency communications~\cite{moldwin2022introduction}. Additionally, increased radiation exposure poses risks to astronauts and those involved in high-altitude flights. Therefore, predicting geomagnetic storms is essential for protecting human life and assets in space and on the ground.

The Disturbance Storm Time (Dst) index measures global geomagnetic activity. Specifically, it reflects Earth's horizontal magnetic field disturbances due to the ring current in the magnetosphere. The Dst index is widely used in space weather to monitor and classify geomagnetic storms according to their value. In this work, we employ symbolic regression to derive data-driven equations that describe the time evolution of the Dst index. To achieve this, we use:
\begin{itemize}
\item Data from the \textbf{NASA OMNIweb database}~\cite{king2005solar}, including the Dst index, solar wind parameters (density, bulk velocity, …), and Interplanetary Magnetic Field (IMF) parameters. This data is obtained from solar wind monitoring spacecraft, including ACE, Wind, IMP-8, and DSCOVR.
 \item  \textbf{\texttt{PySR}, a symbolic regression library} based on evolutionary algorithms~\cite{cranmer2023interpretable}.  With \texttt{PySR}, we can search for optimal mathematical expressions that capture the underlying relationships between dDst/dt and key solar wind parameters, such as the convective electric field, dynamic pressure, magnetic pressure, and Dst itself. By varying the equation complexity, we recover a hierarchy of models, ranging from simple empirical-like formulations to more complex expressions that reveal non-linear dependencies. This approach allows us to systematically compare different equations to describe the temporal evolution of the Dst and assess their physical interpretability.
\end{itemize}

This paper aims to derive data-driven mathematical equations that describe the temporal evolution of the Dst index using symbolic regression. This work uses solar wind, IMF parameters, and Dst historical data from the NASA OMNIweb database. It employs the \texttt{PySR} framework to investigate interpretable relationships between geomagnetic storms and solar wind properties and IMF. The contributions of this work are as follows:
\begin{itemize}
    \item We employ symbolic regression to derive data-driven equations for the Dst temporal evolution (dDst/dt), using solar wind parameters such as the convective electric field, dynamic pressure, magnetic pressure, and the Dst index, obtained from the OMNIweb database.
    \item We explore a hierarchy of models, obtained with \texttt{PySR}, by varying equation complexity as input of \texttt{PySR}. We recover equations with increasing physical accuracy.
    \item We compare the discovered equations with well-established empirical models, such as the Burton-McPherron-Russell~\cite{burton1975empirical} and O’Brien-McPherron models~\cite{o2000forecasting,o2002seasonal}, in accuracy. We show that the best models found with the symbolic regression approach outperform these established models in the cases considered.
    \item We present the advantages of symbolic regression for geomagnetic storm modeling by providing interpretable, closed-form expressions rather than black-box predictions, such as those provided by neural networks.
\end{itemize}

\section{Background and Related Work}
The Dst index measures variations in Earth's geomagnetic field. It is expressed in nanoteslas (nT) as the magnetic fields and primarily reflects the strength of the equatorial ring current, which intensifies during geomagnetic storms. The ring current consists of energetic ions and electrons, which can create a magnetic field in the opposite direction of Earth’s intrinsic geomagnetic field. When the ring current increases, such as during geomagnetic storms, it causes a weakening of the Earth’s field. A negative Dst value corresponds to a reduction in the Earth’s surface magnetic field strength due to the intensification of the ring current. Formally, the Dst index is defined as the horizontal component perturbation on equatorial magnetometers~\cite{sugiura1963hourly}, and its first definition dates back to von Humboldt in the early 18th century~\cite{malin1991humboldt}. The Dst is measured as the longitudinally averaged part of the external field at the geomagnetic dipole equator on the Earth's surface. The World Data Center provides its values for Geomagnetism in Kyoto, and they are also available and tabulated in the NASA OMNIweb database~\cite{king2005solar}, which we use in this work.

One of the most popular definitions of \ul{geomagnetic storm is based on a threshold Dst value, e.g., -50 or -100, below which an event is categorized as a geomagnetic storm}~\cite{borovsky2017dst}. The geomagnetic storms can also be classified depending on their minimum Dst value~\cite{gonzalez1994geomagnetic} as:
\begin{itemize}
\item \textbf{Moderate}: minimum Dst between -50~nT and -100~nT.
\item \textbf{Intense/Great}: minimum Dst between -100~nT and -250~nT.
\item \textbf{Super/Extreme}: minimum Dst below -250~nT.
\end{itemize}

\begin{figure}[h]
    \centering
    \includegraphics[width=0.6\textwidth]{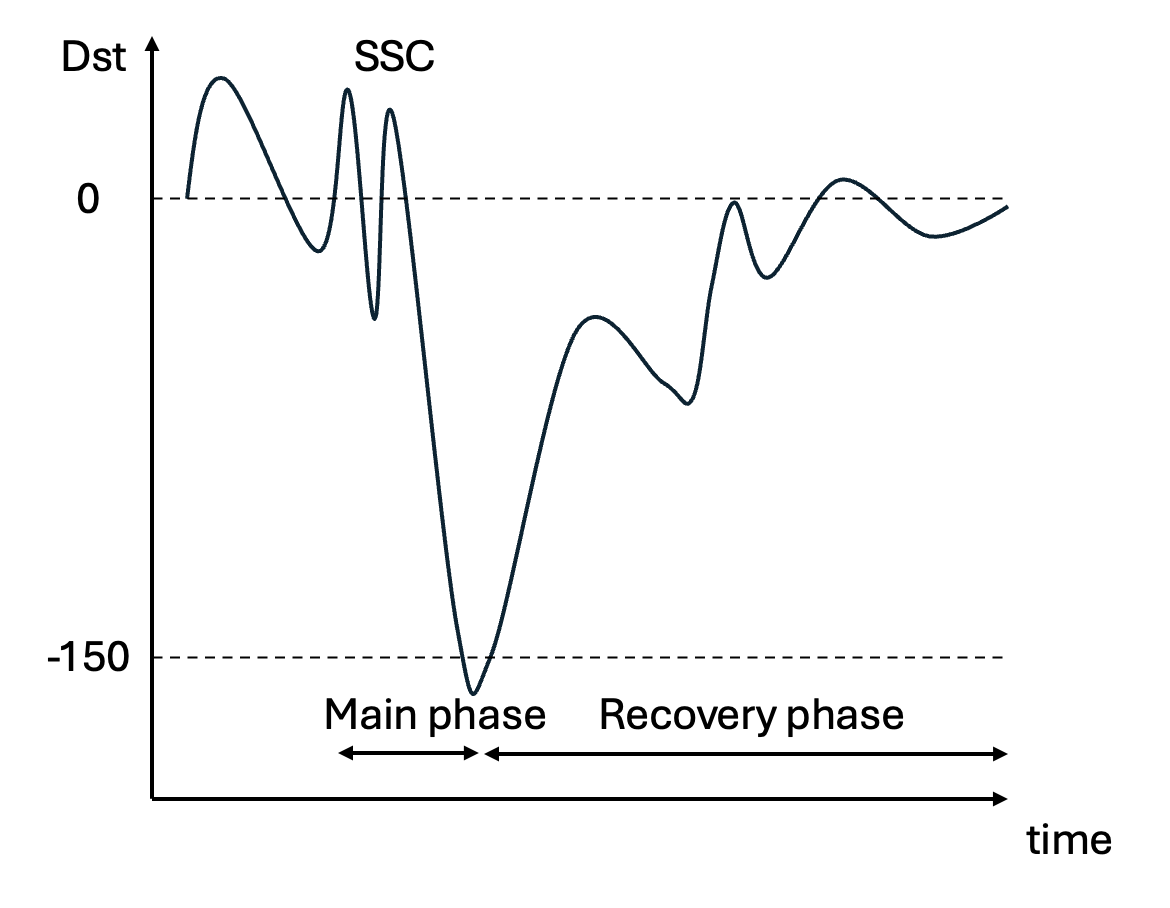}
    \caption{Typical Dst evolution during a geomagnetic storm.}
    \label{fig:Dst1}
\end{figure}
The evolution of the Dst index during a geomagnetic storm follows three phases as depicted in Fig.\ref{fig:Dst1}: 
\begin{itemize}
\item An initial phase with increased Dst, often called \textbf{Sudden Storm Commencement (SSC)}, when an interplanetary shock compresses the magnetosphere, frequently associated with a coronal mass ejection.
\item A \textbf{main phase}, where the Dst index rapidly drops as solar wind-driven particle injection enhances the ring current.
\item A \textbf{recovery phase}, where Earth’s magnetic field returns to its pre-storm state. 
\end{itemize}
The evolution of the Dst index can be compared to a capacitor charging process in an RLC circuit: the magnetosphere can be seen as a capacitor storing energy, and the sudden increase in current is analogous to a transient current increase in an electrical circuit~\cite{cid2005physical}.

The time evolution of the Dst index has been studied using a range of models, varying from first-principles physics-based approaches (such as global MHD models coupled with ring current models~\cite{glocer2013crcm+,jordanova2018specification,lapenta2013swiff}) to empirical models. Due to their simplicity and physical insights, empirical formulations that approximate dDst/dt based on solar wind parameters are among the most widely used. The Burton-McPherron-Russell (BMR) model~\cite{burton1975empirical} is a first and successful example, describing dDst/dt as a balance between solar wind-driven ring current injection and exponential decay due to charge exchange and ionospheric losses. The model uses an input function proportional to the convective electric field, which is linked to the strength of solar wind-magnetosphere coupling and controls the efficiency of energy injection into Earth's magnetosphere. The dynamic pressure of the solar wind also plays an important role by compressing the magnetosphere, modulating the ring current response, and influencing the initial phase of geomagnetic storms. Later improvement, such as the O’Brien-McPherron (OBM) model~\cite{o2000forecasting}, introduced nonlinear decay terms and improved parameterizations to capture storm-time dynamics better. Neural networks have also been explored for predicting dDst/dt, using their ability to model highly nonlinear relationships~\cite{gleisner1996predicting,watanabe2002prediction,gruet2018multiple}. However, while neural networks often achieve high predictive accuracy, they function as black-box models: it is unclear how input variables contribute to the output. AI-driven symbolic regression provides a complementary approach by directly discovering data-driven equations from observational datasets. Unlike neural networks, symbolic regression produces explicit mathematical expressions, allowing for better interpretability and physical insight.

\texttt{PySR} is a symbolic regression framework designed for discovering interpretable mathematical expressions from data using evolutionary search. It is developed primarily in the Julia programming language and provides a Python interface used in this work~\cite{cranmer2023interpretable}. \texttt{PySR} uses evolutionary algorithms: it searches for optimal symbolic expressions by iteratively evolving equations, using operations such as addition, multiplication, and exponentiation, as provided by the user. One of the \texttt{PySR} advantages over other symbolic regression frameworks, such as \texttt{gplearn}~\cite{stephens2019gplearn}, and \texttt{AI Feynman}~\cite{udrescu2020ai}, is its flexible operator set: this includes conditionals (\texttt{if-else}), \texttt{min}, and \texttt{max} functions. These are essential operators for modeling non-linear and threshold-dependent events in the magnetosphere, such as geomagnetic storms. \texttt{gplearn} is a genetic programming-based tool for symbolic regression. It has a customizable function set, but does not natively include \texttt{if-else} logic without user-defined extensions. \texttt{AI Feynman} supports a range of nonlinear and algebraic operations, including polynomial and rational functions. However, its support for conditionals and piecewise functions is more limited compared to \texttt{PySR}. \texttt{AI Feynman} is primarily designed for cases where exact functional relationships exist, such as analytical physics laws. In contrast, \texttt{PySR} is more adaptable to complex, noisy datasets where no exact equation is known a priori.

\section{Methodology}

\subsection{OMNIWeb Dataset and Preprocessing}
The dataset utilized in this study is obtained from the NASA OMNIweb dataset, a collection of near-Earth solar wind, IMF parameters, and geomagnetic indices. Our dataset ranges from January 1, 1995, to May 31, 2021, and includes measurements of solar wind plasma parameters, IMF components, and geomagnetic activity indicators, such as the Dst, at a 1-hour resolution. Our primary target variable is the rate of change of the Dst index, dDst/dt, expressed in nT/hr. Following the examples of previous empirical, physics-based models of the temporal evolution of the Dst, our input variables are shown in Table~\ref{tab:omni_parameters_dst} and include the solar wind speed, the IMF North-South component in GSM coordinates, the solar wind proton density, and the IMF  magnitude. 
\begin{table}[ht]
\centering
\scriptsize
\renewcommand{\arraystretch}{1.3} 
\setlength{\tabcolsep}{4pt}       
\caption{Main physical quantities from the NASA OMNIWeb database included in this study, their units, and potential impact on Dst modeling.}
\label{tab:omni_parameters_dst}
\begin{tabular}{|c|>{\centering\arraybackslash}m{2cm}|>{\raggedright\arraybackslash}m{7.2cm}|}
\hline
\textbf{Quantity} & \textbf{Symbol and Unit} & \textbf{Potential Impact on Dst Modeling} \\ \hline \hline
\rowcolor{lightergray}
Solar Wind Speed & \( V_{SW} \) (km/s) & Controls the convection of solar wind energy toward Earth's magnetosphere, impacting energy injection rates. \\
Interplanetary Magnetic Field North-South Component & \( B_z \) (nT) & Negative \( B_z \) enhances magnetic reconnection efficiency~\cite{thesse1988theoretical} at the dayside magnetopause, intensifying geomagnetic storms. \\
\rowcolor{lightergray}
Solar Wind Density & \( n_{SW}  \) (cm\(^{-3}\)) & Higher solar wind density increases the mass flux of the solar wind, enhancing compression of Earth's magnetosphere. \\
Magnetic Field Magnitude & \( |B| \) (nT) & Represents the strength of the interplanetary magnetic field, influencing pressure balance at the magnetopause. \\
\rowcolor{lightergray}
Solar Wind Temperature & \( T_{SW} \) (K) & It affects the thermal pressure component of the solar wind, modulating solar wind-magnetosphere interaction dynamics. \\ \hline \hline
\end{tabular}
\end{table}

Data pre-processing involves interpolating missing values using a linear scheme, followed by forward and backward filling to ensure continuity. The time derivative of the Dst index is computed using a central finite difference approximation. We then calculate derived quantities, which previous studies have demonstrated to impact the temporal evolution of the Dst. These derived quantities are:
\begin{itemize}
    \item \textbf{Convective Electric Field} ($E_y$), which is defined as $E_y = -V_{\rm SW}\, B_z \times 10^{-3} \text{(mV/m)}$. The magnitude of $E_y$ reflects how strongly the solar wind couples with Earth's magnetosphere. In physics-based models of the Dst, \ul{$E_y$ is used to estimate the energy injection rate into the ring current}.
    \item \textbf{Dynamic Pressure} ($P_{\rm dyn}$), which is calculated from the solar wind proton density and speed $ P_{\rm dyn} = 1.6726 \times 10^{-6} \, n_{SW}  \, V_{\rm SW}^2\text{(nPa)}$, where the pre-factor accounts for the proton mass and necessary unit conversions. The solar wind has two important effects in the context of geomagnetic storms. First, high dynamic pressure leads to a compression of the magnetosphere, e.g., it causes the magnetopause to move inward. Second, strong solar wind shocks with high dynamic pressure can increase ring current formation.
    \item \textbf{Magnetic Pressure} ($P_B$), which is given by $P_B = \frac{B^2}{2\mu_0}$ with $\mu_0 = 4\pi \times 10^{-7}$ being the permeability of free space. In the context of geomagnetic storms, the magnetic pressure is the contribution of the IMF and Earth’s magnetic field to the overall pressure balance in the magnetosphere. The magnetic pressure can affect the magnetopause position by balancing the solar wind pressure.
\end{itemize}
%Several unit conversions are performed in this work. In particular, the derivative dDst/dt is expressed in nanotesla per hour (nT/hr). The convective electric field is given in millivolts per meter (mV/m), and the dynamic pressure is expressed in nanopascals (nPa). 

\subsection{Exploratory Analysis}
\begin{figure}[h]
    \centering
    \includegraphics[width=0.7\textwidth]{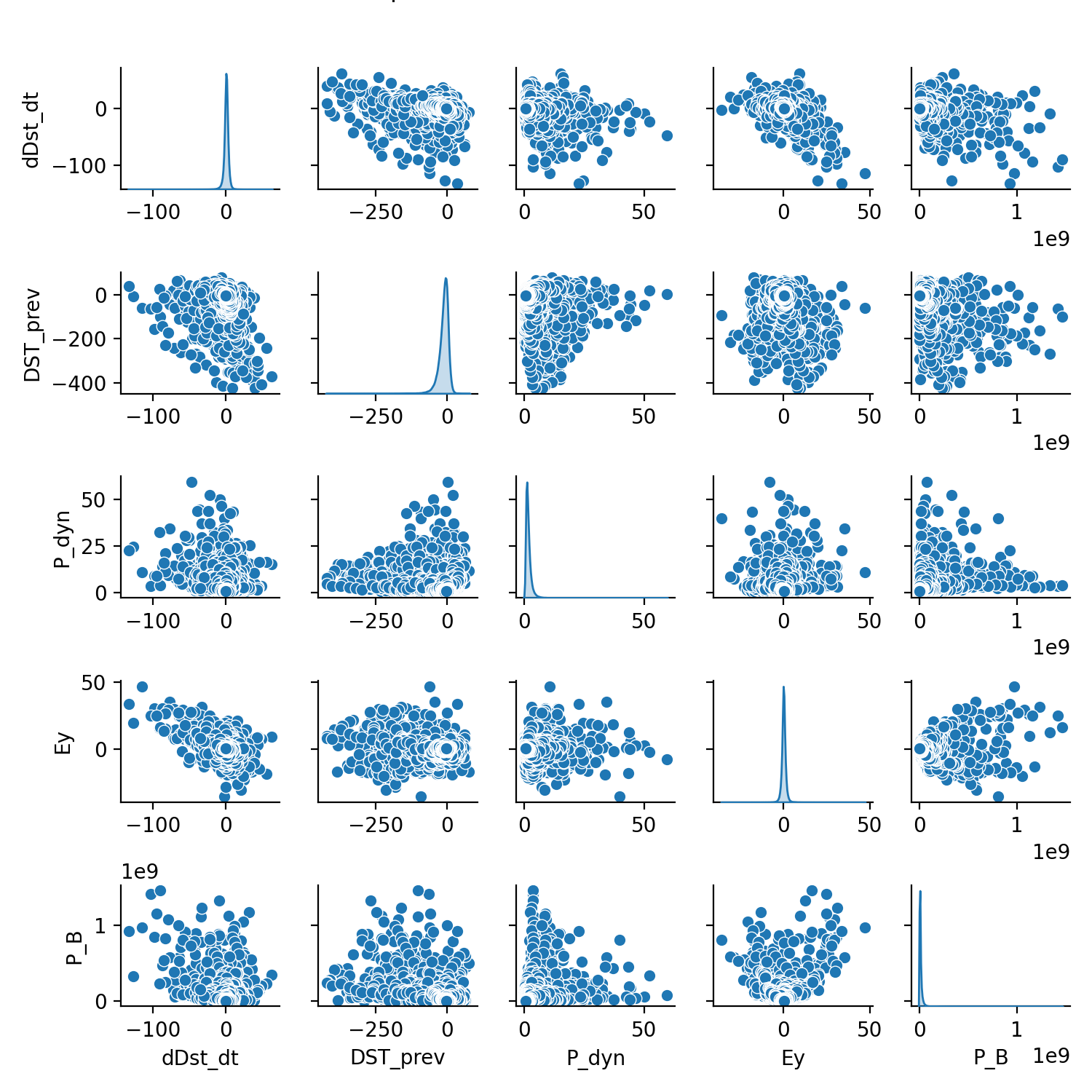}
    \caption{Pairplot of the primary variables considered in this study: $d\text{Dst}/dt$, \texttt{DST\_prev}, $P_{\rm dyn}$, $E_y$, and $P_B$. The diagonal subplots display histograms or density plots for each variable, while the off-diagonal subplots show pairwise relationships.}
    \label{fig:pairplot}
\end{figure}

 We first conduct an exploratory analysis to assess potential relationships among the variables impacting the evolution of the Dst. Fig.~\ref{fig:pairplot} presents a pairplot comparing $d\text{Dst}/dt$, the previous Dst value (\texttt{DST\_prev}), $P_{\rm dyn}$, $E_y$, and $P_B$. In the pairplot, the diagonal entries display the distributions of each variable. The off-diagonal plots show pairwise scatter plots, which show potential correlation or anti-correlation, if the points appear along one line, or non-linear trends. Analyzing Fig.~\ref{fig:pairplot}, we can first focus on distributions along the diagonal of the pairplot. The distribution shows that $d\text{Dst}/dt$ has peaked near zero, indicating that most variations in the Dst index occur in relatively small increments. The distribution of \texttt{DST\_prev} shows that the Dst index peaks on slightly negative values of the Dst with a long tail in the negative Dst direction, which comprises intense and extreme geomagnetic storms. When inspecting the different scatter plots, we can focus on different relationships between the quantities of this study:

\begin{itemize}
    \item \textbf{Relation between dDst/dt and \texttt{DST\_prev}}. The panel in the second row and first column shows the relation of dDst/dt and Dst value at the previous measurement, e.g., an hour before. Overall, the relation is linear with a negative slope, corresponding to Pearson’s anticorrelation of -0.18. We also note that a number of points in the scatter plots are clearly divided but still aligned along a line with a negative slope. \ul{This separation indicates some thresholding phenomenon}, such as when Dst reaches an increased value.
    \item \textbf{Relation between dDst/dt and $P_{\rm dyn}$}. The scatter plot representing the relation of dDst/dt and $P_{\rm dyn}$ does not show any linear behavior (Pearson Correlation: 0.17), hinting at the \ul{presence of a non-linear coupling between the Dst evolution and the dynamic pressure}. In empirical models, this relation typically follows a square root dependence. 
    \item \textbf{Relation between dDst/dt and $E_y$}. When investigating the relation between Dst/dt and the convective electric field (for instance, the panel in the fourth row and first column), we note an overall linear dependence with a negative slope (Pearson correlation: -0.43). We also identify clear outliers for large negative temporal variations of the Dst index: this might indicate the presence of some thresholding event.
    \item \textbf{Relation between Dst/dt and $P_B$}. Similarly to the relation between dDst/dt and $P_{\rm dyn}$, we find a non-linear relationship between the two quantities. However, we observe a larger number of outliers.
\end{itemize}

\begin{tcolorbox}[boxrule=0.5pt, colback=gray!20, colframe=black, sharp corners,]
\textbf{Observation I:} The exploratory analysis shows that \( dDst/dt \) exhibits a strong coupling with the solar wind convective electric field \( E_y \), and nonlinear dependencies on solar wind dynamic pressure. These patterns point to threshold-driven responses in the ring current during geomagnetic storm evolution.
\end{tcolorbox}

\subsection{Symbolic Regression with \texttt{PySR}}
\begin{figure}[ht!]
\centering
\includegraphics[width=0.9\linewidth]{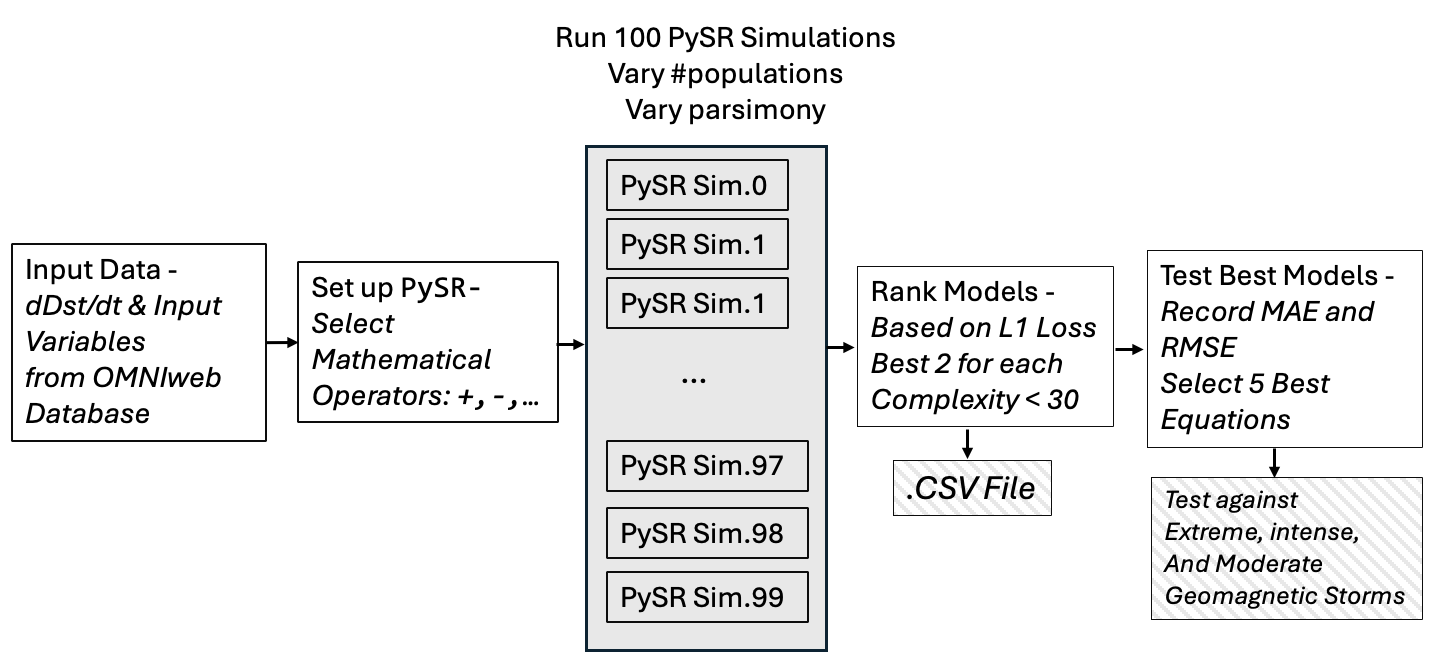}
\caption{A diagram showing the methodology for discovering mathematical models that predict dDst/dt. The data includes data input, symbolic regression with varying hyperparameters, and final model selection based on evaluation metrics.}
\label{fig:method}
\end{figure}

We run \texttt{PySR} to discover mathematical models that capture the relationship between our input variables and $d\text{Dst}/dt$, which is the target. The overall methodology is shown and summarized in Fig.~\ref{fig:method}.
The \texttt{PySR} regression algorithm searches the space of candidate expressions constructed using operators the user selects. In our study, we select the following operators which could be used in our $d\text{Dst}/dt$ model: $+$, $-$, $\times$, $\div$, $\max$, $\min$, $\exp$, $\log$, $\sqrt{\cdot}$, $(\cdot)^2$, and $\operatorname{sign}(\cdot)$. The explored models are evaluated with an L1 loss function, also called Mean Absolute Error (MAE). Since L1 loss is based on absolute differences, it is less sensitive to outliers. 

Two crucial concepts in \texttt{PySR} are the complexity and parsimony of an equation/model. The \texttt{PySR} \ul{complexity of an equation is based on its expression tree}: each mathematical operation, constant, or variable contributes to the total complexity score, potentially with different weights, e.g., trigonometric functions have higher complexity weights than basic operations, such as sum. \texttt{PySR} uses an evolutionary search for symbolic regression and has two main parameters: the number of populations and iterations. The population size determines the number of different candidate solutions that exist in each generation. Increasing the population size improves exploration but also increases the computational cost. The number of iterations defines how many times the evolutionary process updates the population to find better equations. A high number of iterations allows the algorithm to refine solutions further. All the equations discovered have a \texttt{PySR} complexity value. 

\ul{Parsimony is the preference for simpler mathematical expressions} over more complex ones during the model fitting process. \texttt{PySR} implements parsimony through a complexity penalty, which is added to our L1 loss. The level of parsimony can be increased by setting a hyperparameter, which acts as a weight to the complexity penalty in the loss function.

The symbolic regression is performed over $100$ independent \texttt{PySR} runs to account for the stochastic nature of evolutionary algorithms. During each optimization run, hyperparameters such as the parsimony coefficient (randomly selected from $[0.0, 0.9]$) and the population size (randomly chosen between 20 and 120) are varied. The test data set used for the regression test spans from January 1, 1995, to March 31, 2021. 

After the 100 \texttt{PySR} runs, the candidate equations are extracted and filtered based on their loss and complexity metrics, neglecting all equations with complexity greater than 30. Duplicate equations across optimization runs are removed, and the remaining candidates are ranked according to their L1 loss, where lower values indicate a better fit to the test dataset. The final ensemble of candidate models is collected in a CSV file, providing an overview of the discovered equations for dDst/dt. The performance of each candidate equation is evaluated using the  Root Mean Square Error (RMSE) and Mean Absolute Error (MAE), computed from May 1, 2021, to October 1, 2021. The Python code for the symbolic regression, the dataset from the OMNI database, and the discovered equations are available on GitHub\footnote{GitHub repository: \url{https://github.com/smarkidis/Dst-Symbolic-Regression}}.

\section{Physical Model Hierarchy from \texttt{PySR}}
When analyzing the equations generated by \texttt{PySR}, we identify the best equations (with the lowest L1 norm) for different complexities. Table~\ref{tab:dst_models_interpretations} shows how increasing equation complexity increases the mathematical description of dDst/dt and the physical description.

\begin{table}[ht]
\centering
\scriptsize
\renewcommand{\arraystretch}{1.3} 
\setlength{\tabcolsep}{4pt}       
\caption{Equations for dDst/dt with increasing complexity and their corresponding physical interpretation. The progression from a basic exponential decay to a fully coupled model shows how more complex equations introduce additional physics.}
\label{tab:dst_models_interpretations}
\begin{tabular}{|c|>{\centering\arraybackslash}m{7cm}|>{\raggedright\arraybackslash}m{6.5cm}|}
\hline
\textbf{Comp.} & \textbf{Equation} & \textbf{Physical Interpretation} \\ \hline \hline
\rowcolor{lightergray}
3  & \( -0.031\, Dst \) & Basic \ul{exponential decay}. \\ 
5  & \( -0.041\, Dst - Ey \) & Adding \( Ey \) to account for \ul{solar wind forcing}. \\ 
\rowcolor{lightergray}
7  & \( -0.05\, Dst - \max\bigl(Ey,\,-0.16\bigr) \) & Introducing \ul{thresholding} to cap \( Ey \)'s effect, simulating \ul{saturation phenomena}. \\ 
9  & 
\(\begin{gathered}
-0.062\, Dst  -\max\Bigl(-0.062,\; Ey/0.638\Bigr)
\end{gathered}\)
& Rescaling \( Ey \), and \ul{refining the coupling efficiency and threshold limits}. \\ 
 \rowcolor{lightergray} 10 & 
\(\begin{gathered}
-0.057\, Dst  -\max\Bigl(\sqrt{P_{\mathrm{dyn}}}\, Ey,\; -0.098\Bigr)
\end{gathered}\)
& Incorporating \( \sqrt{P_{\mathrm{dyn}}} \) to modulate \( Ey \), to reflect the \ul{influence of solar wind dynamic pressure}. \\ 
12 & 
\(\begin{gathered}
\min\Bigl\{\Bigl[-0.05\, Dst - Ey\Bigr] \times\sqrt{P_{\mathrm{dyn}}},\; -0.055\, Dst\Bigr\}
\end{gathered}\)
& Balancing a dynamic pressure–modulated driver with pure decay, letting the dominant effect prevail. \\ 
\rowcolor{lightergray} 19 & 
\(\begin{gathered}
\Bigl[-0.036(P_{\mathrm{dyn}}+Dst)  - \max(-0.008\,Dst,\,Ey)\Bigr] \times\sqrt{P_{\mathrm{dyn}}+ 1.278} + 0.319
\end{gathered}\)
& Combining \( Dst \) and \( P_{\mathrm{dyn}} \) with refined thresholding and a constant offset, representing a \ul{fully coupled system with background effects}. \\ \hline \hline
\end{tabular}
\end{table}
% CHECK the parentheses in the 19!!!!
With complexity 3, the simplest model captures a basic exponential decay of the Dst index, representing a basic relaxation process. This term corresponds to the decay of the ring current over time due to various loss processes. Note that -1/0.031 corresponds to the decay constant found by the regression (1/0.031 hours $\approx$ 32 hours). 
% if we use 0.031 -> 1/0.031 = 32 hours
The model begins to consider external forcing effects by introducing the solar wind electric field $E_y$ in the equation with complexity 5. In particular, $E_y$  represents the rate at which energy from the solar wind is injected into the Earth's magnetosphere, increasing the ring current. Further \emph{physical} refinements are included in the equations with complexities 7 and 9: these incorporate thresholding nonlinearities, adjust the impact of $Ey$, and simulate saturation effects in the magnetospheric response. Including dynamic pressure $P_{\mathrm{dyn}}$ in complexity 10 modulates the influence of  $E_y$. This reflects the increased energy transfer under varying solar wind conditions. Finally, the most complex model in Table~\ref{tab:dst_models_interpretations} with complexity 19 combines these elements—coupling $DST$ and $P_{\mathrm{dyn}}$, applying thresholding, and including a baseline offset to include terms for modeling quiet times. Note that magnetic pressure is not included in any of the equations discovered by \texttt{PySR}. This suggests that the symbolic regression process identifies the primary geophysical drivers of dDst/dt, in alignment with physics-based models in space physics, where $E_y$ and $P_{\mathrm{dyn}}$ are the dominant contributors to geomagnetic activity.

\begin{tcolorbox}[boxrule=0.5pt, colback=gray!20, colframe=black, sharp corners,]
\textbf{Observation II:} The symbolic regression models progressively incorporate physical processes, from pure Dst exponential decay to solar wind energy injection via \( E_y \) and compression effects through \( P_{\mathrm{dyn}} \). Higher-complexity models capture nonlinear couplings and saturation thresholds.
\end{tcolorbox}

\section{Evaluation and Ranking of Data-Driven Dst/dt Models}
To assess and rank the equations obtained from \texttt{PySR}, we perform a comparative evaluation using a subset of the NASA OMNIweb dataset from May 1, 2021, to October 1, 2021. The tests use 2,000 randomly selected initial conditions. For each valid initial condition, a 48-hour prediction of the Dst index is generated using 11 different equations, ranging in complexity from 25 down to 15.  

The 48‐hour prediction is achieved by iteratively integrating the dDst/dt equation using an Euler-like method. We start from an initial actual Dst value, calculate the dDst/dt at each hourly step by including the current values of $E_y$ and $P_{\mathrm{dyn}}$, rather than relying on their initial values. The performance of each equation is then quantified by computing the RMSE and MAE over these 48-hour intervals, providing an evaluation of the different models.
\begin{figure}[ht!]
\centering
\includegraphics[width=0.8\linewidth]{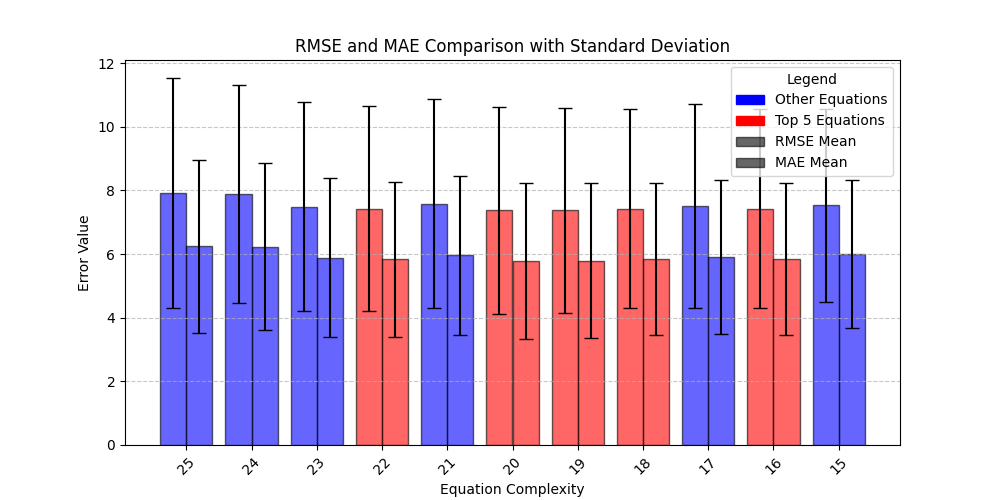}
\caption{RMSE and MAE comparison with standard deviation for data-driven Dst/dt models of different complexities. The error bars indicate the standard deviation across the test dataset. The red bars show the five best-performing models (lowest RMSE and MAE).}
\label{fig:rmse_mae_plot}
\end{figure}
Fig.~\ref{fig:rmse_mae_plot} shows the RMSE and MAE values for various data-driven models, discovered by \texttt{PySR}, categorized by their complexity (on the x-axis). The error bars represent the standard deviation of the different equations. 

Among the models analyzed, the five \texttt{PySR} best-performing models using the test dataset are indicated in red and explicitly presented in Table~\ref{eqson}, which also includes the equations for the established dDst/dt model, BMR, and OBM.
\begin{table}[t]
\centering
\footnotesize
\renewcommand{\arraystretch}{1.3}
\caption{Equations for Dst/dt: BMR, OBM, and data-driven models (DDM) \#1,\#2, \#3, \#4, \#5 with different levels of complexity (C).}
\begin{tabular}{p{2.0cm} >{\raggedright\arraybackslash}p{10.5cm}}
\toprule
\textbf{Model} & \textbf{Equation} \\
\midrule
\midrule
\rowcolor{lightergray} DDM\#1 (C:19)  &
\(\displaystyle 
\Bigl[-0.036\,(P_{dyn}+Dst) - \max(-0.008\,Dst,\, Ey)\Bigr] \sqrt{P_{dyn}+1.278} + 0.319\)
\\[2ex]
DDM\#2 (C:20) &
\(\displaystyle 
\Bigl[-0.042\,(P_{dyn}+Dst) - \max(0.168,\, Ey)\Bigr] \times \sqrt{P_{dyn}+\max(0.097,\, \min(Ey,\, 3.385))} + 0.381\)
\\[2ex]
\rowcolor{lightergray} DDM\#3 (C:18) &
\(\displaystyle 
 \min\Bigl(-0.0443\,Dst,\; (0.621+\sqrt{P_{dyn}})\Bigl[-0.0443\,Dst-\max(Ey,-0.728)\Bigr]\Bigr) + 0.194\)
\\[2ex]
DDM\#4 (C:16) &
\(\displaystyle 
\min\Bigl(-0.0443\,Dst,\; (0.621+\sqrt{P_{dyn}})\Bigl[-0.0443\,Dst - Ey\Bigr]\Bigr) + 0.194\)
\\[2ex]
\rowcolor{lightergray} DDM\#5 (C:22) &
\(\displaystyle 
\min\Bigl(\sqrt{P_{dyn}+1.058}\Bigl[-0.0434\,Dst-Ey\Bigr],\; -0.0434\,Dst\Bigr) + 0.136\,(2.537-0.735\,P_{dyn})\) \\[2ex]
\midrule
\midrule
 BMR  &
\(\displaystyle 
 -0.13\Bigl(Dst - 0.2\,\sqrt{P_{dyn}} + 20.0\Bigr) + 
\begin{cases}
-5.4\,(Ey-0.5) & \text{if } Ey\ge 0.5,\\[1mm]
0 & \text{if } Ey<0.5
\end{cases}\)
\\[2ex]
\rowcolor{lightergray} OBM  &
\(\displaystyle 
 -\frac{1}{\tau}\Bigl(Dst - 0.2\,\sqrt{P_{dyn}} + 20.0\Bigr) + 
\begin{cases}
-5.4\,(Ey-0.5) & \text{if } Ey\ge 0.5,\\[1mm]
0 & \text{if } Ey<0.5
\end{cases}, \quad \quad \quad \quad  \tau = 
\begin{cases}
7.7 & (Ey<0.5)\\[0.5ex]
3.5 & (Ey\ge 0.5)
\end{cases}\)

\\
\bottomrule
\bottomrule
\end{tabular}
\label{eqson}
\end{table}
We use the five best-performing \texttt{PySR} models to predict the Dst evolution of geomagnetic storms and compare them to established models.
\begin{tcolorbox}[boxrule=0.5pt, colback=gray!20, colframe=black, sharp corners,]
\textbf{Observation III:} Increasing complexity does not necessarily improve predictive accuracy for Dst evolution. Models with moderate complexity achieve the best balance between performance and physical interpretability, outperforming complex formulations.
\end{tcolorbox}

\subsection{Performance of Data-Driven Models in Real Geomagnetic Storm Scenarios}
To assess the accuracy of the five best data-driven models, discovered with \texttt{PySR} and presented in Table~\ref{eqson}, in predicting geomagnetic storms, we test five data-driven models against the BMR and OBM models using actual storm events. The performance of these models was evaluated by comparing their predicted Dst with actual Dst values over 72-hour periods. The accuracy of each model was quantified using the RMSE and MAE metrics. 

\begin{figure}[ht!]
    \centering
    \includegraphics[width=0.8\textwidth]{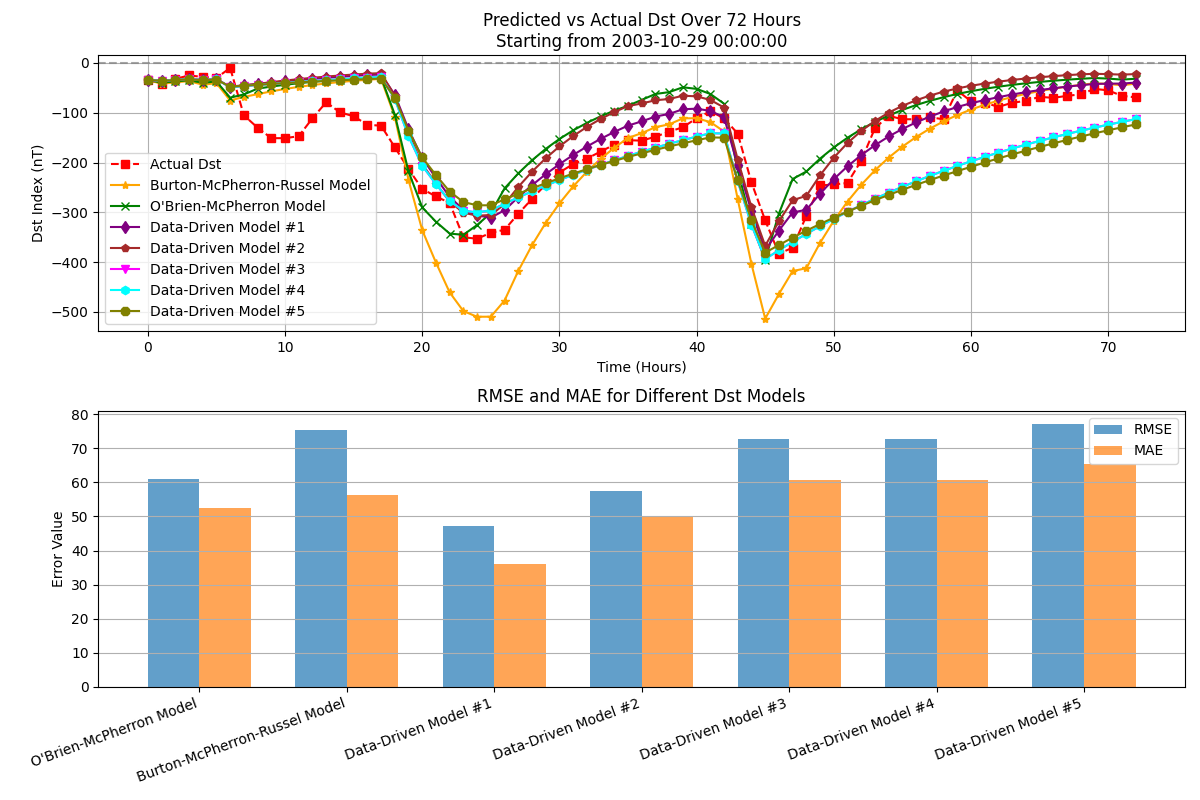}
    \caption{Comparison of Dst predictions for the extreme Halloween Storm (October 29--30, 2003). The top panel shows the predicted vs. actual Dst, while the bottom panel presents RMSE and MAE errors.}
    \label{fig:extreme_event}
\end{figure}

The first event we evaluate is the extreme geomagnetic storm that occurred on 29-30 October 2023. This geomagnetic storm has been widely studied in the literature and goes under the name of the Halloween Storm. Fig.~\ref{fig:extreme_event} shows the Dst prediction for the best \texttt{PySR} models and compares with BMR and OM models on the top panel, and RMSE and MAE in the bottom panel. When investigating the Dst prediction, we note that the BMR model overestimates the lowest Dst value during the geomagnetic storm, and data-driven models \#4 and \#5 do not correctly model the recovery from the second Dst minimum. The models with the lowest MAE errors are, in order, data-driven model \#1 (complexity: 19) and data-driven model \#2 (complexity: 20), followed by the OBM. 
\begin{figure}[ht!]
    \centering
    \includegraphics[width=0.8\textwidth]{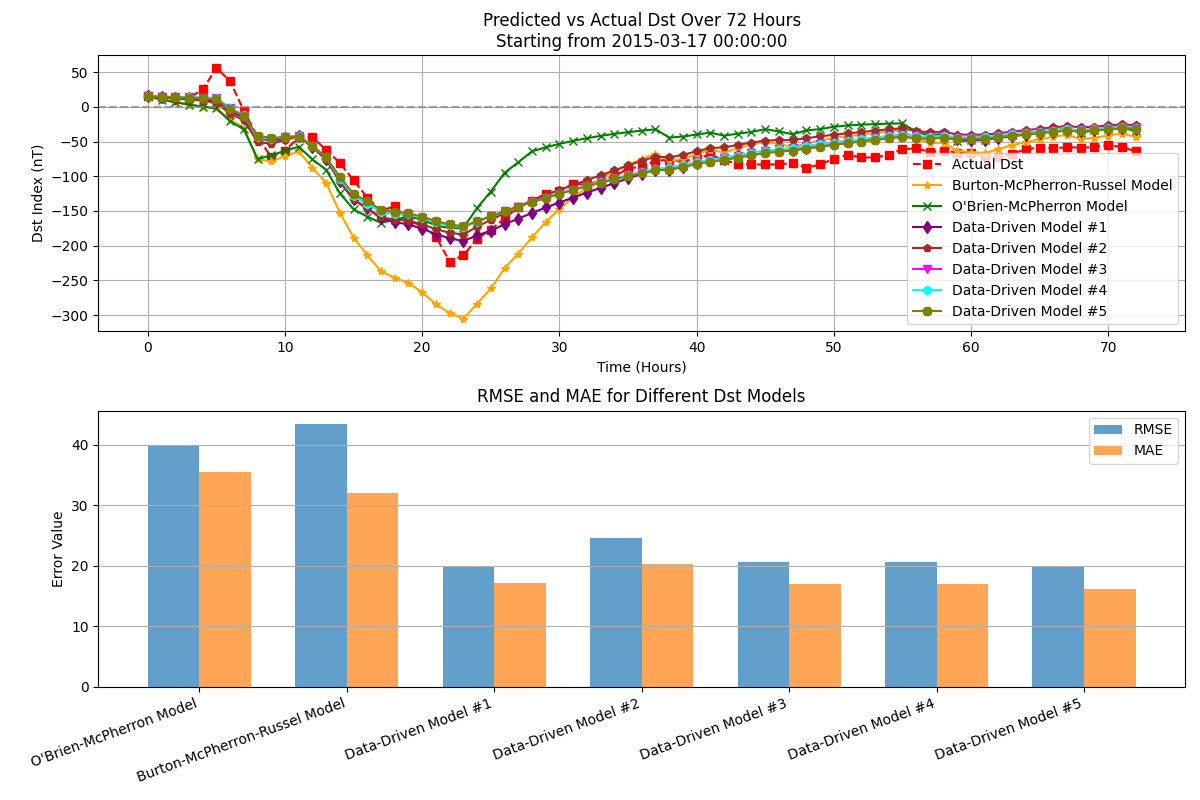}
    \caption{Comparison of Dst predictions for the intense St. Patrick's Day Storm (March 17, 2015) in the top panel and MAE and RMSE errors for different models in the bottom panel. The data-driven models, found by \texttt{PySR}, outperform the BMR and OBM models in terms of MAE and RMSE.}
    \label{fig:stpatricks_storm}
\end{figure}
As the second event, we select an intense/great geomagnetic storm that occurred on March 17, 2015, the so-called St. Patrick’s Day Storm of 2015. The top panel of Fig.~\ref{fig:stpatricks_storm}) shows the Dst prediction and good performance of all the data-driven models compared to the actual Dst. The BMR model underestimates the Dst minimum value as in the previous event, while the OBM does not correctly capture the recovery phase. In this Dst prediction, all the data-driven models outperform the BMR and OBM models.

\begin{figure}[ht!]
    \centering
    \includegraphics[width=0.8\textwidth]{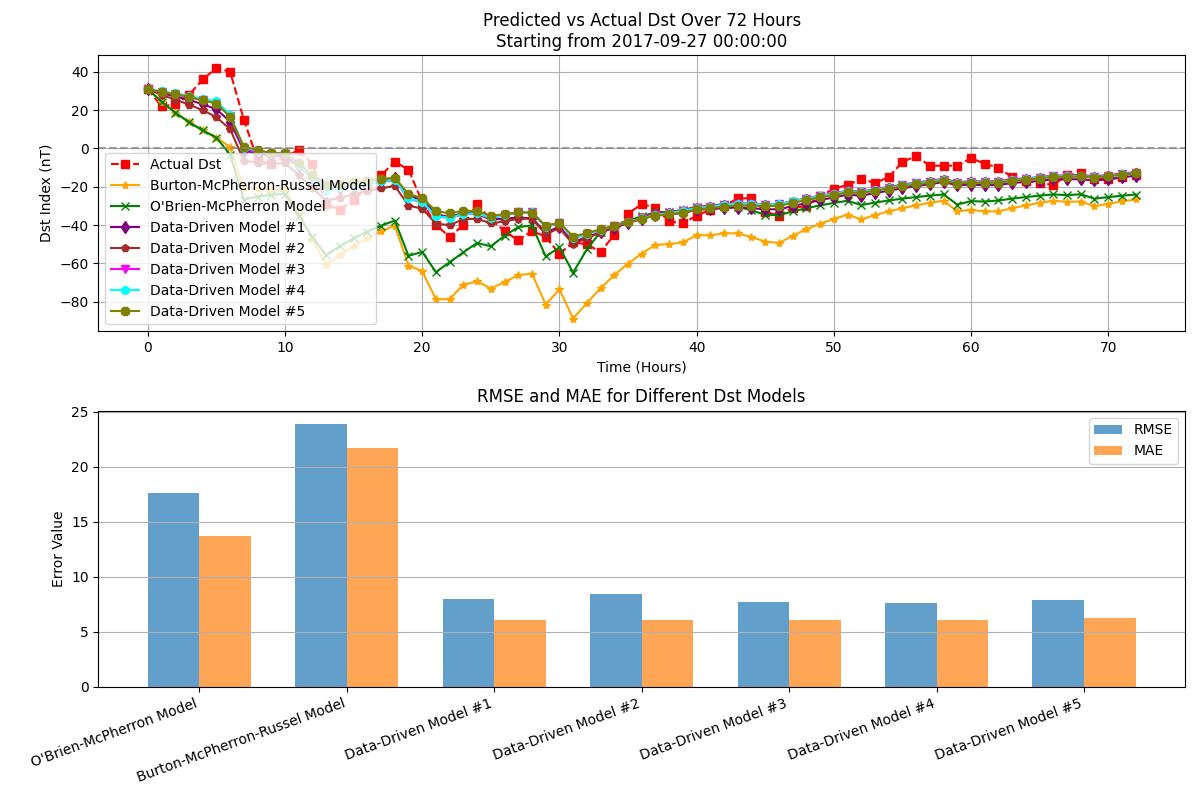}
    \caption{Comparison of Dst predictions for the moderate geomagnetic storm on September 27, 2017, over 72 hours. The data-driven models outperform the semi-empirical BMR and OBM models for moderate geomagnetic storms.}
    \label{fig:moderate_storm}
\end{figure}
Finally we asses the accuracy of different models against a moderate storm occurred on September 27, 2017. Fig.~\ref{fig:moderate_storm}) shows the Dst prediction, along with the RMSE and MAE, for the different models. For this moderate storm, as in the case of intense storms, the data-driven models consistently outperform the BMR and OBM models. This is likely because the number of moderate and intense geomagnetic storms is higher than that of extreme geomagnetic storms in the training dataset. However, in extreme cases, such as the 2003 Halloween event, only two equations discovered by \texttt{PySR} provided better accuracy than the BMR and OBM models.

\begin{tcolorbox}[boxrule=0.5pt, colback=gray!20, colframe=black, sharp corners,]
\textbf{Observation IV:} The symbolic regression models outperform the BMR and OBM empirical models during moderate and intense storms, achieving lower prediction errors for Dst evolution. Only the highest-ranked symbolic models maintain superior performance for extreme events like the 2003 Halloween storm.
\end{tcolorbox}

\section{Conclusion}
This work applied symbolic regression to derive data-driven equations for the temporal evolution of the Dst index, quantifying the evolution of geomagnetic storms. Using NASA OMNIweb data and \texttt{PySR}, we explored models and equations of increasing complexity and compared them with established empirical models for dDst/dt prediction. The methodology used \texttt{PySR} to identify interpretable expressions linking dDst/dt to solar wind parameters, IMF, and the Dst index itself.

The equations discovered by \texttt{PySR} showed increased accuracy compared to traditional empirical models, such as the BMR and OBM models. The best-performing data-driven models, ranked based on RMSE and MAE, captured nonlinear dependencies, including thresholding effects and dynamic pressure modulation. A key aspect is that models discovered by \texttt{PySR} provide high performance while maintaining physical interpretability, unlike other AI-based methods that rely on neural networks. Performance evaluation against real geomagnetic storms, including the 2003 Halloween Storm, the 2015 St. Patrick’s Day Storm, and a moderate storm in 2017, showed that the data-driven equations consistently outperformed BMR and OBM in most cases. However, only the highest-ranked symbolic regression models demonstrated superior accuracy during extreme geomagnetic storms.

\section*{Acknowledgments}
This work is supported by the European Commission, with Automatics in Space Exploration (\href{https://asap-space.eu/}{ASAP}), project no. 101082633. The OMNI data were obtained from the GSFC/SPDF OMNIWeb interface at \url{https://omniweb.gsfc.nasa.gov}.

\newpage

%\bibliographystyle{acm} 
%\bibliography{DstPaperSymbRegArxiv}

\end{document}